\documentclass[
aps,prd,
12pt,
nopreprintnumbers,
showpacs,
eqsecnum,
nofootinbib
]{revtex4-1}

\usepackage[dvipdfmx]{graphicx} 
\usepackage{amssymb}
\usepackage{setspace}
\usepackage{bm}
\usepackage{float} 
\usepackage{here}

\begin{document}

\title{Isothermal spheres from grand partition functions in
nonextensive statistical mechanics}
\author{Nahomi Kan}\email[]{kan@gifu-nct.ac.jp}
\affiliation{National Institute of Technology, Gifu College,
Motosu-shi, Gifu 501-0495, Japan}
\author{Takuya Maki}\email[]{maki@jwcpe.ac.jp}
\affiliation{Japan Women's College of Physical Education,
Setagaya, Tokyo 157-8565, Japan}
\author{Kiyoshi Shiraishi}
\email[]{Deseased 4 November 2024}
\affiliation{
Graduate School of Sciences and Technology for Innovation, 
Yamaguchi University, Yamaguchi-shi, Yamaguchi 753--8512, Japan
}


\begin{abstract}
We analytically study isothermal spheres in the light of nonextensive statistical
mechanics.
The equations for the isothermal spheres are derived from the grand partition
function of the gravitating particle system in the Tsallis statistical mechanics.
The effect of nonextensive statistics appears in relatively dense state,
which appears at the center of the isothermal sphere.
The stability of the isothermal sphere in the general relativistic system is found
to be sensitive to the parameter
$q$ in the Tsallis statistics.
\end{abstract}

\maketitle

\section{Introduction}
\label{sec1}
There has been widespread interest in the equilibrium shape and universal
properties of self-gravitating systems for quite some time \cite{Chandra}. In the
research field of astrophysics, statistical mechanical studies have been conducted
for a long time on the distribution of galaxies and globular clusters \cite{BT}.
Theoretically, gravothermal instability
\cite{LW,Padmanabhan,Padmanabhan2,Chavanis2,Chavanis3} has been analyzed mainly
using the Boltzmann--Gibbs statistical mechanics, and Antonov's analysis of an
isothermal sphere \cite{Antonov} is a famous classic in this field. In the case of
finite temperature, when the density distribution is determined by the equilibrium 
of gravity and pressure gradient,  scales such as the Jeans length
appear in various states. Further, since gravity is a long range force, it follows
that the extensivity of thermodynamical quantities is fundamentally not preserved.

Many studies have been reported on gravitational many-body statistical systems when
Newtonian gravity is extended to general relativity \cite{Chavanis1}. The
properties of the isothermal sphere in general relativity and Ho\v{r}ava gravity
\cite{Horava1,Horava2} were recently clarified by two of the present authors
\cite{KS1}. These have been analyzed based on the statistical mechanics
of Boltzmann--Gibbs.

On the other hand, Tsallis' thermostatistical mechanics \cite{Tsallis} is a
nonextensive theory%
\footnote{For more general statistcs and applications of them, see
\cite{Nojiri:2022aof,Nojiri:2022dkr,Odintsov:2022qnn,Nojiri:2023bom}.} that
includes an additional parameter $q$. It is connected to the canonical
Boltzmann--Gibbs theory in the limit of $q\rightarrow 1$. Tsallis' nonextensive
statistical mechanics has already been used to analyze various statistical systems
including hadronic matter systems, plasma systems, network systems, and systems
with fractal and multifractal structures
\cite{TsallisText,euro}. Nonextensive statistical mechanics has also been actively
applied to systems under the action of gravity, and many papers have been
published on the subject
\cite{TS1,TS2,TS3,Chavanis4,Chavanis5}.
This is because the standard Boltzmann--Gibbs statistics are not considered to
be strictly valid for the system with long-range forces.
Furthermore, there is a motivation to investigate
the relevance of Tsallis statistics for describing celestial systems, since
Plastino and Plastino \cite{PP}, and Taruya and Sakagami \cite{TS1,TS2,TS3}
reported that the equation of state is expressed by the polytrope with the index
$n$ which depends on $q$.  

The theme of the present paper is the exploration of the properties of an
isothermal sphere using a nonextensive statistical mechanical approach. In our
analysis, we use the
grand partition function of the gravitating particle system under the
nonextensive statistics. The technique using the grand partition function is
similar to that in our previous paper \cite{KS1}, except that the statistics are
changed to nonextensive statistics. We start our analysis by introducing a
classical gravitational field, even in the case of the Newtonian gravity,
which is involved in the Hamiltonian. This
allows us to smoothly move on to the consideration of general relativistic
particle systems later. 

This paper is structured as follows. In Section \ref{sec2}, we derive the
approximate analytic equations for a spherically symmetric gravitational field in a
particle system subjected to Newtonian gravity from the grand partition function
in Tsallis' nonextensive statistical mechanics, and discuss the properties of an
isothermal sphere. In particular, we focus on their dependence on the parameter
$q$. The treatment of general relativistic gravitational systems in nonextensive
statistical mechanics is presented in Section \ref{sec3}, and the gravitational
binding energy in the system is studied. 
The final section is devoted to a summary and future prospects.

\section{Grand partition function in the Tsallis statistical mechanics 
for particles in the Newtonian gravity and isothermal spheres}
\label{sec2}

Consider a massive-particle system that is subjected to Newtonian gravity.
Here, we use the coupled classical theory of particles and fields to treat an
isothermal self-gravitating system using nonextensive statistical mechanics, and
investigate the spherically symmetric equilibrium state of the particle system. 

The Newtonian gravitational field equation is a Poisson equation such as
\begin{equation}
\nabla^2\phi(\vec{x})=4\pi G\varrho(\vec{x})\,,
\end{equation}
where $\phi$ is the Newtonian potential and $G$ is
the Newton constant.
The mass density $\varrho$ is here assumed to be that of the gass
consisting of
$n$ particles:
\begin{equation}
\varrho(\vec{x})=m\tilde{\varrho}(\vec{x})=m\sum_{a=1}^n
\delta^3(\vec{x}-\vec{q}_a)\,,
\end{equation}
where $m$ is the common mass of the particles and $q_a$ is the location of
 the $a$th particle.
In this case, the Hamiltonian of the gravitating $n$-particle system is
\begin{equation}
H_n=\int\left[\frac{1}{8\pi
G}(\nabla\phi(\vec{x}))^2+m\tilde{\varrho}(\vec{x})\phi(\vec{x})+
\sum_{a=1}^n\frac{p^2_a}{2m}\delta^3(\vec{x}-\vec{q}_a)\right]d^3\vec{x}\,.
\end{equation}
where $p^a_i$ denotes the momentum of the $a$th particle.

The $n$-particle partition function in Tsallis' statistical mechanics
\cite{Tsallis} is
\begin{equation}
Z_n=\int [D\phi]\int\int\prod_{a=1}^n\frac{d^3\vec{p}_ad^3\vec{q}_a}{(2\pi)^3}
\left[1-\beta(1-q)H_n\right]^{\frac{1}{1-q}}\,,
\label{24}
\end{equation}
where it should be noted that the partition function can be represented by
the path integral with the field variable $\phi$, and
the constant  $\beta$ is the inverse of the temperature, and $q$ is the parameter.
One can see that $\lim_{q\rightarrow 1}(1-\beta (1-q) H)^{1/(1-q)}=e^{-\beta H}$
and thus the limit $q\rightarrow 1$ recovers the partition function of the
Boltzmann--Gibbs.%
\footnote{The $q$-exponential is
frequently defined as $\exp_q X\equiv[1+(1-q)X]^{\frac{1}{1-q}}$, and then 
$\lim_{q\rightarrow 1}\exp_q X=\exp X$.}

The grand partition function $Z_G$ is defined by \cite{GNS}
\begin{equation}
Z_G=\sum_{n=0}^\infty \frac{z^n}{n!}Z_n\,,
\end{equation}
where $z$ is the activity (the fugacity).
Note that nonextensivity makes it impossible to analytically calculate the momentum
integral of each particle separately from the contribution of the other part in the
Hamiltonian. For this reason, we use the integral transformation used by Prato
\cite{Prato} to evaluate the grand partition function. Using Prato's
transformation, we obtain the $n$-particle partition function in Tsallis'
statistical mechanics (\ref{24}):
\begin{equation}
Z_n=\frac{1}{\Gamma\left(\frac{1}{q-1}\right)}\int[D\phi]\int\int\prod_{a=1}^n
\frac{d^3\vec{p}_ad^3\vec{q}_a}{(2\pi)^3}\int_0^\infty
\frac{dt}{t^{\frac{1}{1-q}+1}}\exp\left\{-\left[1-\beta(1-q)H_n\right]t\right\}\,.
\end{equation}
Here, the trace of the particle part of this expression can be rewritten as
\begin{eqnarray}
&
&\int\int\prod_{a=1}^n\frac{d^3\vec{p}_ad^3\vec{q}_a}{(2\pi)^3}\exp\left[-\beta(q-1)t\sum_{a=1}^n
\left(\frac{p^2_a}{2m}+m\phi(q_a)\right)\right]\nonumber \\
&=&\left(\frac{m}{2\pi\beta(q-1)t}\right)^{\frac{3n}{2}}
\int\prod_{a=1}^nd^3\vec{q}_a
\exp\left[-\beta(q-1)t\sum_{a=1}^n
m\phi(q_a)\right]\nonumber \\
&=&\left[\int d^3\vec{x}\left(\frac{m}{2\pi\beta(q-1)t}\right)^{\frac{3}{2}}
e^{-\beta(q-1)tm\phi(\vec{x})}\right]^n\equiv\left[G(t)\right]^n
\equiv\left[\int g(t) d^3\vec{x}\right]^n\,.
\label{gt}
\end{eqnarray}
Therefore, the grand partition function for $q>1$ is obtained as
\begin{equation}
Z_G=\int[D\phi]\int_0^\infty
\frac{dt}{\Gamma\left(\frac{1}{q-1}\right)t^{\frac{1}{1-q}+1}}\exp\left\{-\left[1-\beta(1-q)\int\frac{1}{8\pi
G}(\nabla\phi)^2 d^3\vec{x}\right]t +z G(t)
\right\}\,.
\end{equation}
The integral associated with the Prato transformation can be evaluated using 
the method of steepest descents \cite{Migdal} (the
saddle point method).%
\footnote{Incidentally, the integral 
\[
\frac{1}{\Gamma(\frac{1}{q-1})}\int_0^\infty
\frac{dt}{t^{\frac{1}{1-q}+1}}\exp[-\beta(q-1)t H_n]
\]
(where $\beta H_n$ is independent of $t$)
is evaluated by the stationary point method
$(\int_0^\infty e^{-f(t)}dt\approx
\sqrt{\frac{2\pi}{|f''(t_1)|}}e^{-f(t_1)}$, where $t_1$ satisfies
$f'(t_1)=0)$
 and results in
\[
\frac{\sqrt{2\pi}e^{-\frac{2-q}{q-1}}
\left(\frac{2-q}{q-1}\right)^{\frac{2-q}{q-1}-\frac{1}{2}}}{\Gamma\left(
\frac{2-q}{q-1}\right)}\left[1-\beta(1-q)H_n\right]^{\frac{1}{1-q}}\,.
\]}
We assume $z\ll 1$ and estimate the integral by the stationary point method up to
the first order of $z$. Later, small $z$ will be related to the low central
density of the isothermal sphere. 
Then, we obtain
\begin{equation}
Z_G=\frac{\sqrt{2\pi}e^{-\frac{2-q}{q-1}}
\left(\frac{2-q}{q-1}\right)^{\frac{2-q}{q-1}-\frac{1}{2}}}{\Gamma\left(
\frac{2-q}{q-1}\right)}\int[D\phi]
\left[1-\beta(1-q)H_{eff}\right]^{\frac{1}{1-q}}\,,
\end{equation}
where
\begin{equation}
H_{eff}=\int d^3\vec{x}\left\{
\frac{1}{8\pi G}(\nabla\phi)^2-z\left[\frac{2-q}{\beta\hat{N}}g(t_0)
+\frac{1}{\beta}g'(t_0)+\frac{\hat{N}}{2\beta(q-1)}g''(t_0)
\right]\right\}\,,
\label{eqg}
\end{equation}
with
\begin{equation}
t_0=\frac{1}{q-1}\hat{N}\,,\quad
\hat{N}=(2-q)\left(1+\beta(q-1)\int\frac{1}{8\pi
G}(\nabla\phi)^2d^3\vec{x}\right)^{-1}\,.
\end{equation}
Substituting $g(t)$ defined in (\ref{gt}) into the expression, we find
\begin{eqnarray}
& &H_{eff}=\int d^3\vec{x}\left\{
\frac{1}{8\pi G}(\nabla\phi)^2\right.\nonumber \\
& &\qquad\qquad\left.-\frac{z e^{-\beta\hat{N}m\phi}}{\beta\hat{N}}
\left[\frac{m}{2\pi\beta\hat{N}}\right]^{3/2}\left[1+
\frac{q-1}{2}\left(-\frac{5}{4}+\beta\hat{N}m\phi
+\beta^2\hat{N}^2m^2\phi^2\right)\right]\right\}\,.
\label{effH}
\end{eqnarray}
Although the integral transform is expressed by the specific contour integration
on the complex plane for $q<1$ \cite{Prato}, the representation of the effective
Hamiltonian is the same as (\ref{effH}) because we use the method of steepest
descents and the integrand is common for the cases with $q>1$ and $q<1$.

Here, we can easily check that it reproduces the result with the Boltzmann--Gibbs
statistical mechanics \cite{VSC1,VSC2}, in the limit of
$q\rightarrow 1$. The previous result of the Liouville type was
obtained by applying the Gaussian integration and perform Stratonovich--Hubbard
transformation. We should also notice that the factor $\hat{N}$ appears only in
the combination $\beta\hat{N}$. Thus, we can recognize that the factor
$\hat{N}$ is absorbed in $\beta$ as  ``renormalization'' in a certain sense.
The equation of motion can be derived
from the variation principle on the effective Hamiltonian.
Note that, in the limit of $q\rightarrow 1$, the equation of motion for $\phi$
becomes a Poisson-type equation
\begin{equation}
\nabla^2\phi=4\pi G\rho_0 e^{-\beta
m\phi}\,,
\end{equation}
where $\nabla^2$ denotes the Laplacian in a flat space and we set $\rho_0\equiv
zm\left(\frac{m}{2\pi\beta}\right)^{3/2}$, since
$\hat{N}\rightarrow 1$ when $q\rightarrow 1$. This equation has already been known
for the Newtonian isothermal gas
\cite{Chandra,BT,Antonov,LW,Padmanabhan,Padmanabhan2,Chavanis2,Chavanis3}.

Now, we consider the case with spherical symmetry.
Then we find the following second-order differential equation:
\begin{equation}
\frac{1}{x^2}\frac{d}{dx}\left(x^2\frac{dy}{dx}\right)=
\frac{1+\frac{q-1}{2}\left(-\frac{9}{4}-y+y^2\right)}%
{1+\frac{q-1}{2}\left(-\frac{9}{4}-y_0+y_0^2\right)}e^{-(y-y_0)}\equiv n(y)\,,
\label{xy}
\end{equation}
where the radial coordinate $r$ and the gravitational potential $\phi$ have been
rescaled as
\begin{equation}
x\equiv\sqrt{4\pi G\rho_c \beta\hat{N}m}\,r\,,
\quad
y\equiv \beta\hat{N}m\phi\,,
\end{equation}
with
\begin{equation}
\rho_c\equiv zm
\left[\frac{m}{2\pi\beta\hat{N}}\right]^{3/2}\left[1+\frac{q-1}{2}
\left(-\frac{9}{4}-y_0+y_0^2\right)\right]e^{-y_0}\,.
\end{equation}
Here, the boundary conditions are $y(0)=y_0$ and $y'(0)=0$, where the prime
$({}')$ denotes the derivative in terms of $x$.
Obviously, the invariance under the arbitrary choice of $y_0$ in (\ref{xy})
is found only in the case of $q=1$, i.e., the case with the authodox isothermal
sphere in the Boltzmann--Gibbs statistical mechanics
\cite{Chandra,BT,Antonov,LW,Padmanabhan,Padmanabhan2,Chavanis2,Chavanis3}. Thus,
we find that the nonextensivity breaks this scale invariance under
$y\rightarrow y+\ell$ and $x\rightarrow x e^{\ell/2}$. As we will see
later, the scale invariance is broken also in the general relativistic isothermal
sphere, even in the case with $q=1$ \cite{KS1}. That is, the properties of the
isothermal sphere also depend on $y_0$, or equivalently, the central value of the
density of gass.

The differential equation (\ref{xy}) can be decomposed the following two
first-order equations, 
\begin{equation}
y'(x)=\frac{\mu(x)}{x^2}\,,\quad \mu'(x)=
n(y)x^2\,.
\end{equation}
Then, we define two functions as in the previous studies
\cite{Chandra,Padmanabhan}:
\begin{equation}
v\equiv\frac{\mu(x)}{x}=xy'(x)\,,\quad u\equiv
\frac{n(y)x^3}{\mu(x)}=\frac{n(y)x}{y'(x)}=\frac{n(y)x^2}{v}\,.
\label{Nuv}
\end{equation}
\begin{figure}[H]
\centering
\includegraphics[width=50mm]{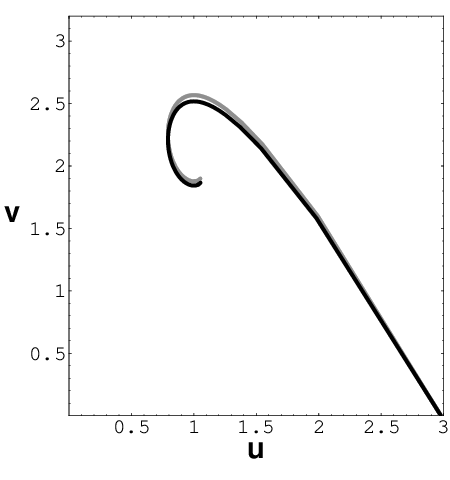}\quad
\centering
\includegraphics[width=50mm]{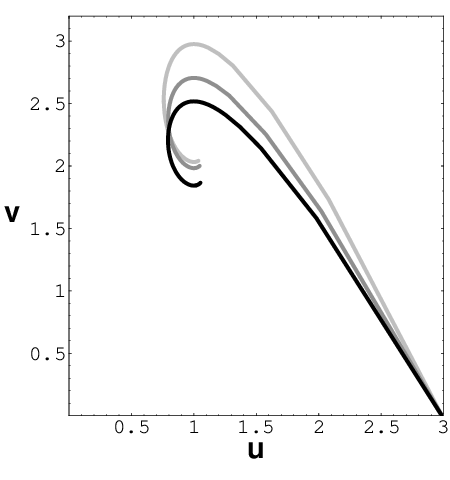}\quad
\centering
\includegraphics[width=50mm]{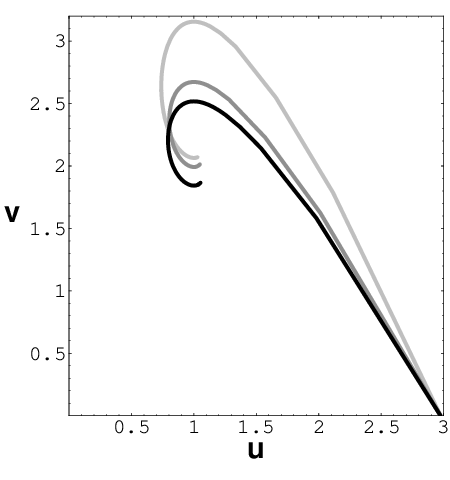}\\
\hspace{5mm}
(a) \hspace{47mm} (b) \hspace{47mm} (c)
\caption{
Plots of spirals in the $(u,v)$ plane for Newtonian isothermal spheres.
}
\label{fig1}
\end{figure}


In Fig.~\ref{fig1}, we show the solutions for various conditions in the $(u, v)$
plane. The black curves indicate the case with $q=1$, and the grey curves
correspond to the case with $q=1.01$, while lighter grey curves indicate the
case with $q=1.1$, in three figures $(a)$,$(b)$, and $(c)$. The spirals become
larger according to larger values of $q$. The curves in the case with low central
density
$y_0=100$ are given in Fig.~\ref{fig1}(a). The relatively high-density cases with
$y_0=10$ [Fig.~\ref{fig1}(b)] and $y_0=6$ [Fig.~\ref{fig1}(c)] exhibit similar
characteristics.
All the curves start at the point $(u,v)=(3, 0)$, which represents the center of
the isothermal sphere ($x=0$), and they approach the fixed point $(u,
v)\approx(1,2)$, which corresponds to $x\rightarrow\infty$.

The behavior of curves near $(u, v)=(3,0)$ is found to be
\begin{equation}
u=3+b v\,\quad (v\ll 1)\,,
\end{equation}
where
\begin{equation}
b=-\frac{3\left[1+\frac{q-1}{8}(-5-12y_0+4y_0^2)\right]}
{5\left[1+\frac{q-1}{8}(-9-4y_0+4y_0^2)\right]}
\,.
\end{equation}
Taruya and Sakagami \cite{TS1} advocated that the gass of massive particles
governed by the Tsallis statistical mechanics obeys the polytropic equation of
state.  
For the polytropic sphere \cite{TS1,Chavanis5}, 
\begin{equation}
b=-\frac{3}{5}\frac{n}{n+1}\,,
\end{equation}
where $n$ is the polytrope index (the pressure $P$ is proportional to
$\rho^{\frac{n+1}{n}}$, where $\rho$ is the density).
Therefore, in our present analysis, the effective polytrope index in the center of
the sphere can be expressed by
\begin{equation}
n=\frac{1}{y_0-1/2}\left(
\frac{1}{q-1}+\frac{-5-12y_0+4y_0^2}{8}\right)\,,
\end{equation}
which depends on $q$ as well as $y_0$, but is independent of the activity $z$.
This result does not coincide with the one in \cite{PP,TS1} or the other
reports by Taruya and Sakagami. Of course, the Boltzmann--Gibbs limit, $q=1$ gives
$n=\infty$, which corresponds to the case of the isothermal sphere.
We also notice that $n=\infty$ is obtained in the low temperature
(equivalently, low density) limit, which is attained by $y_0\rightarrow\infty$.

In our analysis, the effects of nonextensivity appear at relatively high
densities. Thus, the three spirals appear almost overlapping in Fig.~\ref{fig1}(a).
Since the density of gass decreases in the asymptotic region $x\gg 1$, all the
spirals in Fig.~\ref{fig1} approach the fixed point $(u, v)\approx (1, 2)$,
which indicates that the equation of state in the asymptotic region corresponds to
$n=\infty$.
In all of these cases, we find the asymptotic behavior $\rho\propto 1/r^2$,
because it turns out that  $v\approx 2$ and $u\approx 1$ means $n\approx 2/x^2$.
Note that the current approximation could not hold up to significant high
densities, since we adopted $z\ll 1$ in the approximation scheme in evaluating
the effective Hamiltonian. For the same reason, we study only on the case with
$q>1$ in the present analysis, otherwise ($q<1$) the central density becomes
apparently negative even for large
$y_0$ under the present approximation.

There are many arguments about stability even in the case of ordinary statistical
mechanics. In the present paper, the stability will be discussed in the next
section, after the consideration of the isothermal sphere in general relativistic
systems.

\section{Grand partition function in the Tsallis statistical mechanics 
for particles in the Einstein gravity and isothermal spheres}
\label{sec3}
In this section, we study  isothermal spheres in the Einstein gravity,
from the viewpoint of Tsallis' statistical mechanics.
The Hamiltonian formalism is adopted for the system in the present analysis.
We start with the ADM line element \cite{ADM},
\begin{equation}
ds^2=-N^2dt^2+\tilde{g}_{ij}(dx^i+N^idt)(dx^j+N^jdt)\,,
\label{ADMm}
\end{equation}
where $N$ is the lapse function, $N^i$ ($i, j=1, 2, 3$) is the shift vector, and
$\tilde{g}_{ij}$ is  the spatial metric.
The Hamiltonian for the gravitational field is then written in the form
\begin{equation}
H=\int d^3x\,\sqrt{\tilde{g}}\left(N\mathcal{H}+N^i\mathcal{H}_i\right)\,,
\end{equation}
where $\tilde{g}$ is the determinant of $\tilde{g}_{ij}$.
Here, the Hamiltonian constraint $\mathcal{H}$ and the momentum constraint
$\mathcal{H}_i$ are given by
\begin{equation}
16\pi G\, \mathcal{H}=\frac{1}{\sqrt{\tilde{g}}}\left[
\pi^{ij}\pi_{ij}-\frac{1}{2}(\pi^i_i)^2\right]-\tilde{R}
\,,\quad
16\pi G\,\mathcal{H}_i=-2\nabla^j\pi_{ij}\,,
\end{equation}
with $\pi^{ij}$ being the conjugate momentum of $\tilde{g}_{ij}$, and $\tilde{R}$
 being the scalar curvature constructed from
$\tilde{g}_{ij}$, respectively. Here, $\nabla_i$ is the three-dimensional
covariant derivative.

We express the Hamiltonian of the relativistic $n$-particle system in the
spacetime as 
\begin{equation}
H_n=N\sum_{a=1}^n\sqrt{\tilde{g}^{ij}p^a_ip^a_j+m^2}\,,
\end{equation}
where $m$ is the common mass of the particles and $p^a_i$ denotes the momentum of
the $a$th particle located at $q_a$.

The treatment of the grand partition function is the
same as in the case with the Newtonian gravity.
We should notice the following relation and the definition of $g_{R}(t)$;
\begin{eqnarray}
&
&\int\int\prod_{a=1}^n\frac{d^3\vec{p}_ad^3\vec{q}_a}{(2\pi)^3}\exp\left[-\beta(q-1)t\sum_{a=1}^n
\left(N(\vec{q}_a)\sqrt{\tilde{g}^{ij}p^a_ip^a_j+m^2}\right)\right]\nonumber \\
&=&\int\prod_{a=1}^nd^3\vec{q}_a\sqrt{\tilde{g}}\,\frac{m^3}{2\pi^2}
\frac{K_2(\beta(q-1)t mN(\vec{q}_a))}{\beta(q-1)
tmN(\vec{q}_a)}\nonumber \\
&=&\left[\int d^3\vec{x}\sqrt{\tilde{g}}\,\frac{m^3}{2\pi^2}
\frac{K_2(\beta(q-1)t mN(\vec{x}))}{\beta(q-1)
tmN(\vec{x})}\right]^n=\left[G_{R}(t)\right]^n=\left[\int g_{R}(t)
d^3\vec{x}\right]^n\,,
\end{eqnarray}
where the special function $K_\nu(z)$ is the
modified Bessel function of the second kind.
The grand partition function can thus be expressed as
\begin{equation}
Z_G=\frac{1}{\Gamma\left(\frac{1}{q-1}\right)}\int[DN][D\tilde{g}_{ij}]\int_0^\infty
\frac{dt}{t^{\frac{1}{1-q}+1}}\exp\left\{-\left[1+\beta N(1-q)\int\frac{1}{16\pi
G}\tilde{R} \sqrt{\tilde{g}} d^3\vec{x}\right]t +z G_{R}(t)
\right\}\,.
\end{equation}
We have assumed that the
shift vector vanishes for nonrotating bodies, and the
integration over conjugate momentum
$\pi^{ij}$ is omitted. In other words, the graviton degrees of freedom are
decoupled from the statistical system.

As in the previous section, we evaluate the integration over $t$ by the method of
steepest descents. Replacing $g(t)$ by $g_R(t)$ in the second half of (\ref{eqg}),
and replacing the kinetic term of $\phi$ in the first half of  (\ref{eqg}) by
$-\frac{1}{16\pi G}N\tilde{R}$, we obtain
\begin{eqnarray}
& &H_{eff}=\int d^3\vec{x}\sqrt{\tilde{g}}\left\{
-\frac{1}{16\pi G}N\tilde{R}\right.\nonumber \\
& &\left.-\frac{zm^3}{2\pi^2\hat{\beta}}
\left[\frac{K_2(\hat{\beta}Nm)}{\hat{\beta}Nm}+
\frac{q-1}{2}\left({K_1(\hat{\beta}Nm)}+
(4+\hat{\beta}^2N^2m^2)\frac{K_2(\hat{\beta}Nm)}{\hat{\beta}Nm}\right)\right]\right\}\,,
\end{eqnarray}
where
\begin{equation}
\hat{\beta}\equiv\beta\hat{N}\,,\quad
\hat{N}=(2-q)\left[1+\beta N(1-q)\int\frac{1}{16\pi
G}\tilde{R} \sqrt{\tilde{g}} d^3\vec{x}\right]^{-1}\,.
\end{equation}
It is noteworthy that $\hat\beta$ appears
 in the combination $\hat\beta N=\hat\beta\sqrt{-g_{00}}$, as advocated by Tolman
\cite{Tolman}.

We can derive equations for the static equilibrium configuration,
which are obtained by the
variation of the effective Hamiltonian. One can obtain the following classical
equations of motion from the variational principle:
\begin{equation}
\tilde{R}=16\pi G\rho(y)\,,
\end{equation}
where
\begin{equation}
\rho(y)=\frac{zm^4}{2\pi^2y}
\left[K_1(y)+\frac{3}{y}K_2(y)+
\frac{q-1}{2}\left(2y\,K_0(y)+(7+y^2)K_1(y)+
12\frac{K_2(y)}{y}\right)\right]\,,
\label{NN}
\end{equation}
and
\begin{equation}
N\left(\tilde{R}_{ij}-\frac{1}{2}\tilde{R}\tilde{g}_{ij}
\right)-\nabla_i\nabla_jN+\nabla^2N\tilde{g}_{ij}
=8\pi G NP(y)\tilde{g}_{ij}\,,\label{EE}
\end{equation}
where
\begin{equation}
P(y)=\frac{zm^4}{2\pi^2y}\left[\frac{K_2(y)}{y}+
\frac{q-1}{2}\left(y K_2(y)+K_3(y)\right)\right]\,.
\label{PP}
\end{equation}
In the above expressions,
\begin{equation}
y\equiv\hat{\beta}Nm\,,
\end{equation}
$\tilde{R}_{ij}$  is the Ricci
tensor constructed from
$\tilde{g}_{ij}$, and $\nabla^2\equiv\tilde{g}^{ij}\nabla_i\nabla_j$.

Now, we discuss the case with spherical symmetry.
Then, we set the metric as
\begin{equation}
ds^2=-N^2(r)\,dt^2+\frac{dr^2}{1-\frac{2GM(r)}{r}}+r^2(d\theta^2+\sin^2\theta\,
d\varphi^2)\,,
\label{metric}
\end{equation}
where  the function $M(r)$ 
describes the mass inside the sphere with radius $r$.%
\footnote{Since the equation from the variation of $N$ gives $M'$, $M$ is
proportional to the volume integral of the $00$ component of the energy-momentum
tensor, which appears in the right-hand side of the Einstein equation.}
Substituting the metric (\ref{metric}), the equations of motion are reduced to
\cite{KS1}
\begin{equation}
\frac{1}{r^2}\frac{dM(r)}{dr}
=4\pi \rho(y)\,, \quad\frac{1}{r}\left[1-\frac{2GM(r)}{r}\right]\frac{dN(r)}{dr}
-\frac{GM(r)}{r^3}N(r)=4\pi G N(r) P(y)\,,
\end{equation}
where $\rho(y)$ and $P(y)$ are defined by (\ref{NN}) and (\ref{PP}) with
$N\rightarrow N(r)$. In order to simplify the equations further, we rescale the
variables,
\begin{equation}
x\equiv\sqrt{4\pi G \rho_c}\, r\,,\quad
y\equiv\hat{\beta} mN\,,\quad
\tilde{M}\equiv\sqrt{4\pi G \rho_c}GM\,,
\end{equation}
with
\begin{equation}
\rho_c\equiv\rho(y_0)\,,\quad y_0\equiv y(0).
\label{rho}
\end{equation}
Then, the equations for $\tilde{M}(x)$ and $y(x)$ read
\begin{equation}
\frac{1}{x^2}\tilde{M}'=\frac{\rho(y)}{\rho_c}\,,\label{eex1}
\quad\frac{1}{x}\left(1-\frac{2\tilde{M}}{x}\right)y'
-\frac{\tilde{M}}{x^3}\,y
=\frac{y P(y)}{\rho_c}\,,
\label{eex2}
\end{equation}
where the prime (${}'$) means the derivative with respect to $x$.
We must find solutions satisfying the boundary conditions
$\tilde{M}(0)=0$ and $y(0)=y_0$.

Now, we define two functions \cite{KS1}:
\begin{equation}
u\equiv \frac{d\ln M(r)}{d\ln
r}=x\frac{\tilde{M}'(x)}{\tilde{M}(x)}\,,
\quad
v\equiv \hat{\beta} m\frac{dN(r)}{d\ln r}=x y'(x)\,.
\end{equation}
It was shown that $u$ and $v$ coincide with the variables in (\ref{Nuv})
in the Newtonian limit \cite{KS1}.

\begin{figure}[ht]
\centering
\includegraphics[width=5cm]{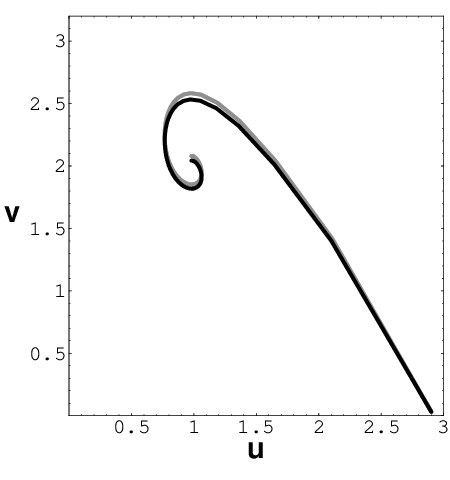}\quad
\includegraphics[width=5cm]{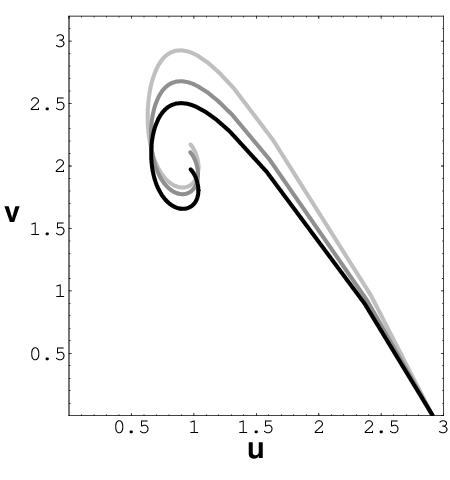}\quad
\includegraphics[width=5cm]{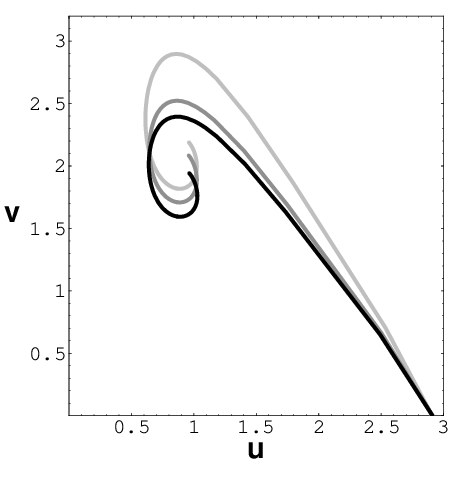}\\
(a) \hspace{47mm} (b) \hspace{46mm} (c)
\caption{Plots of spirals in the $(u,v)$ plane for general relativistic isothermal
spheres.\\
 (a)
$y_0=100$, (b)
$y_0=10$, and (c) $y_0=6$. From the inner spiral to the outer spiral, $q=1$,
$q=1.01$, and $q=1.1$ in each plot.}
\label{fig2}
\end{figure}

In Fig.~\ref{fig2}, we show the general relativistic solutions for various initial
conditions in the
$(u, v)$ plane.
The three figures correspond to the boundary conditions $y_0=100$,
$10$, and $6$, respectively. The correspondences of grey-scales of curves are the
same as in Fig.~\ref{fig1}. All the curves start at the point $(u, v)=(3,0)$, which
represents the center of the isothermal sphere. The fixed point $(u,v)\approx
(1,2)$ is almost unchanged.

In the relativistic case, the slope coefficient $b$, in the behavior of curves
 $u=3+b v\,\quad (v\ll 1)$  near $(u, v)=(3,0)$, is less than $-\frac{3}{5}$.
Therefore, the polytropic interpretation
of equation of state is not so effective in the center of isothermal spheres.

While the behaviors of the spirals in the $(u,v)$ plane
and the density profiles have moderate dependence on the central density and the
parameter $q$,
the parameter dependence of stability is very complicated as we will show below.
Therefore, in the present paper, we only discuss the stability by considering the
ratio of the sum of the mass of constituent particles and the mass of the
isothermal sphere in the region of the fixed radius. The analyses using various
known methods are left for future studies.

Because the isothermal spheres in our model have the similar asymptotic density
profile as the Newtonian one, we consider the finite spherical box to define the
mass of the object
\cite{Chandra,BT,Antonov,LW,Padmanabhan,Padmanabhan2,Chavanis2,Chavanis3}. We
consider the region inside the sphere with radius $r$.

Here, we consider the ratio of the mass of the particles and the total mass
$M$, which includes gravitational binding energy,  $mN_p/M$, expressed as
\cite{KS1}
\begin{equation}
\frac{m N_p}{M}=\frac{1}{\tilde{M}(x)}\int_0^x
\frac{m n_p(y(x')){x'}^2}{\rho_c\sqrt{1-
\frac{2\tilde{M}(x')}{x'}}}dx'\,,
\end{equation}
where the particle number density is given by
\begin{equation}
n_p(y)=\hat{\beta}NP=\frac{zm^3}{2\pi^2}
\left[\frac{K_2(y)}{y}+
\frac{q-1}{2}\left({K_1(y)}+
(4+y^2)\frac{K_2(y)}{y}\right)\right]\,,
\end{equation}
which is derived from the effective Hamiltonian.%

\begin{figure}[ht]
\centering
\includegraphics[width=5cm]{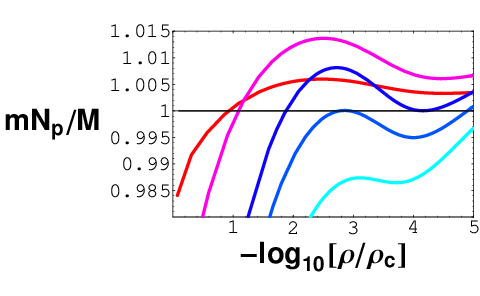}\quad
\includegraphics[width=5cm]{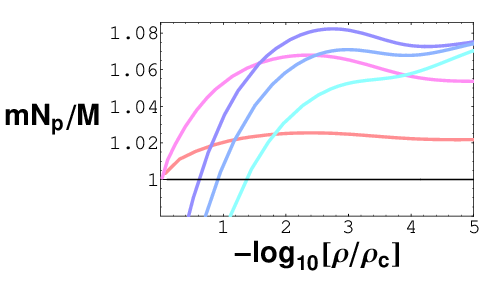}\quad
\includegraphics[width=5cm]{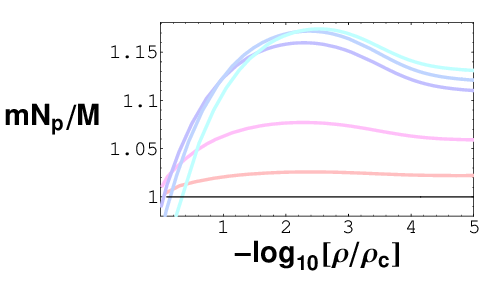}\\
\hspace{1.5cm}(A) \hspace{4.5cm} (B) \hspace{4.5cm} (C)
\caption{The ratio $mN_p/M$ is plotted. (A) $q=1$, (B) $q=1.01$, and
(C) $q=1.1$. In each plot, the red curve represents $y_0=100$,
the magenta curve is for $y_0=30$,
the blue curve is for $y_0=10$,
the pale blue curve is for $y_0=7.8$, and
the cyan curve is for $y_0=6$.}
\label{fig3}
\end{figure}

Fig. \ref{fig3} shows the ratio $mN_p/M$ versus $-\log_{10}[\rho/\rho_c]$.
Since the ratio $\rho(r)/\rho_c$ monotonically decreases with $r$ (or $x$), as we
have seen in this section, the radius of the spherical box becomes larger from
left to right in the horizontal axis. Notice that, apart from some exceptions, the
characteristic feature of the plots lies inside the finite region by selecting the
scale of the axis. 

Since
$N_p=\int n_p
\sqrt{\tilde{g}}d^3x$, the ratio
$mN_p/M>1$ implies positive binding energy $m N_p-M$, so the isothermal sphere
is expected to be energetically stable in this case.
We should also consider the maximum point of the ratio. On the right-hand side of
the maximum, the increase of the radius of the sphere reduces the amount of 
binding energy per mass. Thus, there is a possibility that some smaller
bodies produced by fission will be more stable than a single body.

In Fig.~\ref{fig3}(A), we show the general relativistic case in the
Boltzmann--Gibbs statistics $(q=0)$ \cite{KS1}. If the central density is
sufficiently large (i.e., the gas is rather relativistic in the vicinity of the
center), the stable configuration disappears according to the above-mentioned
criteria. Its critical value is
$y_0\approx 7.8$. However, the change of the curve is complicated for $y_0> 10$.
For $y_0=100$, the maximum is located around $-\log_{10}[\rho(r)/\rho_c]\approx
2.5$. This is consistent with the known stability criterion for the Newtonian
isothermal sphere $-\log_{10}[\rho(r)/\rho_c]<2.85$
\cite{Antonov,LW,Padmanabhan,Padmanabhan2,Chavanis2,Chavanis3}.

The cases of the nonextensive statistics are also complicated.
The behaviors of the curves in Figs.~\ref{fig3}(B) (for $q=1.01$) and
\ref{fig3}(C) (for
$q=1.1$) drastically change if $y_0$ is larger than about $10$.
Thus, the almost nonrelativistic gas sphere is stable when the radius is
relatively small. 
Another important feature one can observe in Fig.~\ref{fig3} is that 
even the relatively high-central-density (small $y_0$) spheres may possess a stable
radius  for $q>1$.

\begin{figure}[ht]
\centering
\includegraphics
{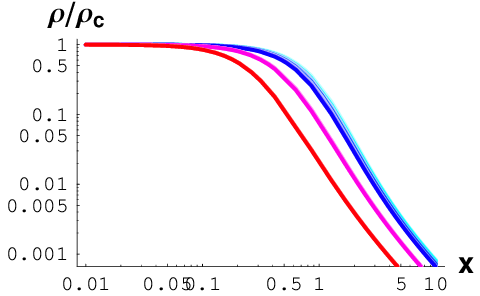}
\caption{Density profiles $\rho(x)/\rho_c$ of general relativistic
isothermal spheres. The colors of curves are exactly the same as those assigned in
Fig.~\ref{fig3}. Note that the curves corresponding to different values of $q$
almost overlap.}
\label{fig4}
\end{figure}

Fig.~\ref{fig4} shows the density profiles $\rho(x)/\rho_c$ as functions of $x$.
The parameters $y_0$ and $q$ are all the same as those in Fig.~\ref{fig3}.
Note that the curves corresponding to different $q$ are nearly overlapping.
In all of these cases, we find the asymptotic behavior $\rho\propto 1/x^2$,
similarly to that of the Newtonian isothermal sphere.

\section{Summary and prospects}
\label{sec6}

Using the grand partition function, we investigate the spherically
symmetric equilibrium state of an isothermal self-gravitating system from the
Tsallis' nonextensive statistical mechanical treatment of the system.
At the central region, where the density is relatively high, the equation of state
can be expressed by the effective polytropic index $n(<\infty)$, which  however
differs from the previously reported result \cite{PP,TS1}. We also considered such
isothermal spheres in curved space.  For
$q>1$, the gravitational stability is attained for wider range of central
densities. The density profiles of isothermal spheres depends on the central
density, but are almost independent of the value of $q$.
In future work, it will be possible to analyze
the fluctuations in this system using field theory techniques.

In our present approach, we only consider the case with $q\ge 1$,
since the central density becomes negative even for low density in the present
analysis. This is due to the approximation which holds only for $z\ll 1$. However,
if it is not the case, the equations of motion becomes highly nonlinear 
integro-differential equations, for both Newtonian and relativistic cases.
Further investigation of the case with $q<1$ may be needed, in relation to
the general treatment of the gravitating system, such as a cluster of galaxies
\cite{HH}.  Some analytic treatment of such cases will be one of the future
challenges.

Since we deal with the grand canonical formalism, we can also
take the mergers  and splits of particles (i.e. the change and transfer in the
particle number) into consideration. In such a case, even spherically symmetric
states may have a layer structure.
On the other hand, extension to cases with a strong gravitational field as a
background field can be considered. This consideration leads to a model of the
system including black holes. 
As we treat the isothermal system, we wish to take
the temperature as the Hawking temperature in the presence of a black hole. To
incorporate such a back ground effect, we must extend the formulation to the case
with fully curved spacetime and inclusion of (Hawking) radiation (and other
possible fields).

Finally, we consider that the study of gravitating systems governed by  quantum
statistical mechanics with nonextensivity and
more general statistics
\cite{Nojiri:2022aof,Nojiri:2022dkr,Odintsov:2022qnn,Nojiri:2023bom}. are much
interesting. The contribution of quantum fields as a correction to the stable
conditions will be an important subject to study in future works.


\acknowledgments
One of the authors (KS) would like to thank A.~Taruya for discussion.
We would like to thank M.~Takeuchi for reading this manuscript.
We would like to thank S.~D.~Odintsov for valuable information on general
statistics and their work. 
Also TM would like to thank A.Sugamoto, S.Ansoldi, the University of Trieste and ICTP for giving TM this research opportunity.

\bibliographystyle{apsrev4-1}


\end{document}